\documentclass[aps,prl,twocolumn]{revtex4-2}
\usepackage[utf8]{inputenc}
\usepackage[T1]{fontenc}

\usepackage{amsmath}
\usepackage{amsfonts}
\usepackage{amssymb}
\usepackage{dsfont}

\usepackage[english]{babel}

\usepackage{physics}

\usepackage{hyphenat}
\usepackage{hyperref}
\usepackage{xcolor}
\usepackage{xspace}
\usepackage{graphicx}

\usepackage[normalem]{ulem}

\begin{document}

\title{Quantum annealing for lattice models with competing long-range interactions}
\author{Jan Alexander Koziol}
\affiliation{Department of Physics, Staudtstra{\ss}e 7, Friedrich-Alexander-Universit\"at Erlangen-N\"urnberg (FAU), D-91058 Erlangen, Germany}
\author{Kai Phillip Schmidt}
\affiliation{Department of Physics, Staudtstra{\ss}e 7, Friedrich-Alexander-Universit\"at Erlangen-N\"urnberg (FAU), D-91058 Erlangen, Germany}

\begin{abstract}
We use superconducting qubit quantum annealing devices to determine the ground state of Ising models with algebraically decaying competing long-range interactions in the thermodynamic limit.
This is enabled by a unit-cell-based optimization scheme, in which the finite optimizations on each unit cell are performed using commercial quantum annealing hardware.
To demonstrate the capabilities of the approach, we choose three exemplary problems relevant for other quantum simulation platforms and material science: 
(i) the calculation of devil’s staircases of magnetization plateaux of the long-range Ising model in a longitudinal field on the triangular lattice, motivated by atomic and molecular quantum simulators;
(ii) the evaluation of the ground state of the same model on the Kagomé lattice in the absence of a field, motivated by artificial spin ice metamaterials;
(iii) the study of models with additional few-nearest-neighbor interactions relevant for frustrated Ising compounds with potential long-range interactions.
The approach discussed in this work provides a useful and realistic application of existing quantum annealing technology, applicable across many research areas in which lattice problems with resummable long-range interactions are relevant.
\end{abstract}

\maketitle

\begin{figure*}[t]
  \centering
  \includegraphics[width=\textwidth]{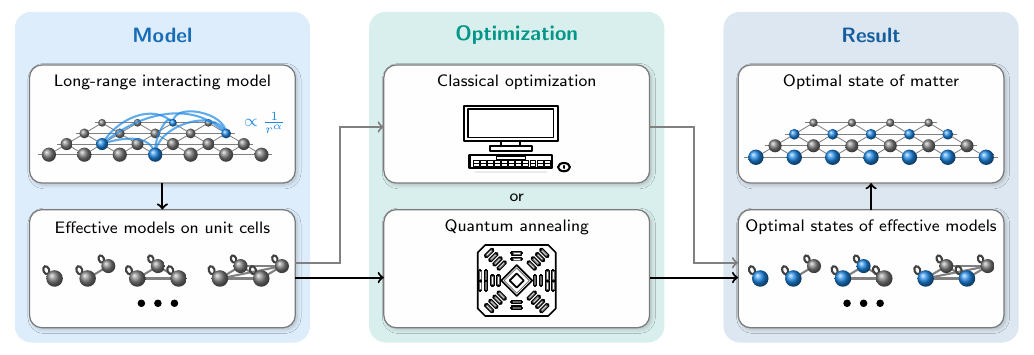}
  \caption{Schematic overview of the conceptual steps in the unit-cell-based optimization scheme (UCBOS). \textit{Left:} The given long-range interacting model is converted into effective models on unit cells. \textit{Middle:} The energy optimization can be done by quantum annealing or classical algorithms. \textit{Right:} The optimal states of the effective models result in the optimal state of matter of the investigated long-range interacting model in the themodynamic limit.}
  \label{fig:overview}
\end{figure*}

%%% Introduction
In recent years, lattice models with algebraically decaying long-range interactions became highly relevant due to their realization in a variety of analogue quantum simulation platforms based on atomic or molecular quantum optics systems \cite{Browaeys2020,Monroe2021,Chomaz2022,Defenu2023} and condensed matter systems \cite{Chioar2016,Skjrv2019,Regan2020,Li2021,Huang2021,Nuckolls2024,Yadav2025}.
Furthermore, quantum annealing or quantum adiabatic optimization \cite{Finnila1994,Kadowaki1998,Farhi2001,Das2008,Johnson2011,Hauke2020,Yarkoni2022,Rajak2022}, the solution of classical optimization problems using an adiabatic ground-state evolution of a quantum-mechanical Hamiltonian \cite{Lucas2014}, is one of the most mature \cite{Hauke2020,Yarkoni2022,Rajak2022} and commercialized \cite{Hauke2020,Yarkoni2022,Rajak2022} applications of the current noisy intermediate-scale quantum (NISQ) era \cite{Preskill2018}.
In this work we demonstrate the successful application of state-of-the-art commercially available on-chip superconducting qubit quantum annealing devices \cite{Johnson2011,Bunyk2014,Boothby2020,King2020,McGeoch2021,McGeoch2022} to determine ground states of classical lattice problems with long-range interactions in the thermodynamic limit.
We implement a unit-cell-based optimization scheme (UCBOS) \cite{Koziol2023,Koziol2024,Koziol2025} where the optimization is run on the Advantage\texttrademark\ system quantum annealing device \cite{Boothby2020,King2020,McGeoch2021,McGeoch2022} of the company D-Wave Systems Incorporated (D-Wave).
Our work therefore represents the timely connection between available quantum hardware and relevant systems with long-range interactions and applications across various experimental platforms in quantum optics and condensed matter physics. 
A schematic overview of our approach is given in Fig.~\ref{fig:overview}.
We show (i) how to investigate all-to-all connected lattice problems with algebraically decaying long-range interactions on finite-connectivity quantum annealing devices; (ii) relevant applications of quantum annealing for the improved understanding of many other cutting edge quantum simulation platforms and material science; (iii) a runtime advantage over current classical binary optimization algorithms for the relevant problem sizes for the UCBOS.

%%% Long-range Ising model
\section{The long-range Ising model}
A paradigmatic workhorse model of systems with long-range interactions is the long-range Ising model (LRIM) in a longitudinal field that occurs naturally in several artificial quantum simulation platforms, e.\,g., in the atomic limit of ultracold dipolar atoms or molecules in optical lattices \cite{Chomaz2022}, Rydberg atoms in the limit of small Rabi frequencies \cite{Browaeys2020}, trapped ions with no effective transverse field \cite{Monroe2021}, or artifical spin ice \cite{Chioar2016,Skjrv2019}. We define the Hamiltonian of the antiferromagnetic LRIM in a longitudinal field with competing algebraically decaying long-range interactions on an arbitrary lattice as
\begin{align}
		\label{eq:LRIM}
		H = h\sum_i\sigma_i^z + \frac{J}{2}\sum_{i\neq j} \frac{1}{|\vec r_i-\vec r_j|^{\alpha}} \sigma_i^z\sigma_j^z
\end{align}
with the Pauli matrices $\sigma_i^z$ describing classical two-level degrees-of-freedom at lattice sites $\vec r_i$.
The amplitude of the Ising coupling is parametrized by $J>0$, the longitudinal field by $h\in\mathds{R}$ and the decay of the long-range interaction by $\alpha$.
Note, for the application of the UCBOS in this work, we require $\alpha$ to be larger than the spatial dimension $d$ of the system.

%%% Unit-cell-based ground-state optimization
\section{Unit-cell-based ground-state optimization}

To determine the ground state of LRIMs \eqref{eq:LRIM} the authors developed a UCBOS in preceeding works \cite{Koziol2023,Koziol2024,Koziol2025}.
The aim of this method is to find ground states in the thermodynamic limit without truncation of the long-range interactions and with minimal bias from the simulation geometries.
For illustration and notation purposes we discuss the topic using the LRIM on two-dimensional lattices.
Note that generalizations to other interaction types and higher dimensional lattices are straightforward.
The key observation enabling this scheme is that one can determine the energy of the LRIM of a periodic state with a $K$-site unit cell and translational vectors $\vec T_1$ and $\vec T_2$ in the thermodynamic limit by considering solely an effective model with appropriately resummed couplings on a unit cell of the state
\begin{align}
	\label{eq:EffectiveModel}
	\frac{1}{2}\sum_{i\neq j}\frac{J}{|\vec r_i -\vec r_j|^\alpha}\sigma_i^z\sigma_j^z= \frac{1}{2}\sum_{i,j=0}^K \bar J^{\alpha}_{i,j}\sigma_i^z\sigma_j^z
\end{align}
with the resummed couplings $\bar J^{\alpha}_{i,j}$ defined as
\begin{align}
		\bar J^{\alpha}_{i,j} & = J \sum_{k=-\infty}^{\infty}\sum_{l=-\infty}^{\infty} \frac{(1-\delta_{i,j}\delta_{l,0}\delta_{k,0})}{|\vec r_i -\vec r_j + l\vec T_1 - k \vec T_2|^\alpha}\\
						  &= J \ \zeta_{\Lambda(\vec T_1, \vec T_2), \alpha}(\vec r_i - \vec r_j, \vec 0) \ .
\end{align}
Here $\delta_{i,j}$ is the Kronecker delta, $\Lambda(\vec T_1, \vec T_2)$ the lattice spanned by $\vec T_1$ and $\vec T_2$, and $\zeta_{\Lambda(\vec T_1, \vec T_2), \alpha}(\vec x, \vec y)$ the Epstein $\zeta$ function \cite{Epstein1903,Epstein1906,Buchheit2021,Buchheit2022,Buchheit2024Code,Buchheit2024Code1}.
We will now invert this reasoning to formulate a scheme for the systematic ground-state search in the thermodynamic limit.
First, we create all relevant and feasible unit cells of interest and determine the effective models with resummed couplings~\eqref{eq:EffectiveModel} for each of the cells.
Then, we determine optimal patterns for the finite problems on the unit cells using a discrete optimization algorithm.
Since the energies are considered in the thermodynamic limit, we know that the optimal state on each unit cell is the optimal state for the thermodynamic limit, among all states that fit the given unit cell.
The final step is to compare the energies of the optimal states between the unit cells, to find the candidate for the overall optimal state.
This approach is limited by two factors: (i) the size of the considered unit cells, (ii) the capabilities of the optimization algorithm on the unit cells.
In preceeding works \cite{Koziol2023,Koziol2024,Koziol2025} using the UCBOS, classical stochastic searches -- with either naive local discrete ``greedy'' \cite{Koziol2023} optimizations, or continuous variable \cite{Koziol2024,Koziol2025} algorithms -- were used.
Although this seems sufficient for the considered unit-cell sizes, there are more involved and potent classical binary optimization approaches available \cite{Kirkpatrick1983,Hukushima1996,Hamze2004,Selby2014,Gogate2004,Zhu2015}.
Here, we will replace the classical binary optimization by a quantum annealing protocol on the D-Wave Advantage\texttrademark\ system and use the quantum device as a hardware accelerator for the Ising optimization problems. 

%%% Quantum annealing
\section{Quantum annealing}
Quantum annealing \cite{Finnila1994,Kadowaki1998,Farhi2001,Das2008,Johnson2011,Hauke2020,Yarkoni2022,Rajak2022} is a computational concept to find the global minimum of an optimization problem by adiabatically evolving a quantum system from an initial simple Hamiltonian to a final Hamiltonian encoding the optimization objective.
The D-Wave quantum annealing devices are purpose-build quantum simulators for quadratic unconstrained binary optimization problems which are equivalent to generalized Ising problems \cite{Matsubara1956}.
The dynamics of the devices can be described by the time-dependent Hamiltonian
\begin{align}
	H = -\frac{A(s)}{2}\underbrace{\left(\sum_i\sigma_i^x\right)}_{H_{\mathrm{I}}} + \frac{B(s)}{2}\underbrace{\left(\sum_i h_i \sigma_i^z + \sum_{i>j} J_{i,j} \sigma_i^z \sigma_j^z\right)}_{H_{\mathrm{F}}}
\end{align}
with $\sigma_i^{x/z}$ being Pauli matrices operating on the qubit $i$, $J_{i,j}$ the coupling strength between qubit $i$ and $j$, and the longitudinal fields $h_i$ called local biases.
The time $s\in[0,1]$ dependent parameters $A(s)$ and $B(s)$ parametrize the tuning of the initial Hamiltonian $H_{\mathrm{I}}$ at the beginning to the final Hamiltonian $H_{\mathrm{F}}$ at the end of the annealing run.
The annealing protocol needs to be chosen in a way to keep the system in its ground state along the entire path.
If successful, the final state realizes the ground-state of $H_{\mathrm{F}}$ and is therefore a global minimum of the encoded optimization problem.
In the D-Wave Advantage\texttrademark\ system hardware ``at least 5,000 qubit'' \cite{McGeoch2022} are placed on a chip and coupled in a Pegasus graph \cite{Boothby2020}.
The Pegasus graph has a finite local connectivity with a qubit degree of 15 and a naive all-to-all connected subgraph of four qubits \cite{Boothby2020}.
Therefore, to embed Ising optimization problems with higher connectivity, the use of chains is required \cite{Klymko2013,Boothby2015,Boothby2020}.
In a chain, multiple hardware qubits are coupled to realize a single logical qubit \cite{Klymko2013,Boothby2015,Boothby2020}.
Efficient algorithms for minor embedding \footnote{The term minor embedding refers to the embedding of a graph $P$ into a graph $F$ where ``$P$ can be obtained from $F$ by a series of edge and site deletions and edge contractions\cite{Klymko2013}''. An edge contraction removes an edge from a graph and merges the two vertices that were previously joined.} of all-to-all connected problem graphs (cliques) to the Pegasus qubit fabric are available \cite{Klymko2013,Boothby2015,Boothby2020}.

\section{Demonstrations}
We choose three paradigmatic features of the antiferromagnetic LRIM to demonstrate and gauge the capabilities of the UCBOS accelerated by the quantum annealing.
First, we determine the devil's staircase of magnetization plateaux of the LRIM in a longitudinal field.
Second, we evaluate the ground state of the LRIM on the Kagomé lattice in the absence of a field.
This constitutes a relevant and debated topic motivated by artificial spin ice \cite{Chioar2016,Skjrv2019}.
Third, we modify the long-range interaction by additional few-nearest-neighbor interactions to demonstrate the applicability for Ising quantum materials with potential long-range interactions \cite{Yadav2025}.

\begin{figure}
	\centering
	\includegraphics[]{"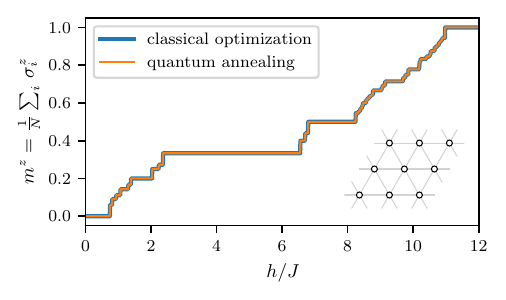"}
	\caption{Devil's staircase of ground-state-magnetization plateaux of the dipolar ($\alpha=3$) antiferromagnetic LRIM in a longitudinal field on the triangular lattice calculated using the UCBOS with classical naive greedy optimization (blue line) and quantum annealing on the D-Wave Advantage\texttrademark\ system (orange line). The magnetizations are calculated on a grid of $10^{-2}$ in $h/J$.}
	\label{fig:staircase}
\end{figure}

We present the devil's staircase of ground-state magnetization plateaux of the dipolar ($\alpha=3$) antiferromagnetic LRIM in a longitudinal field on the triangular lattice in Fig.~\ref{fig:staircase}.
We expect an infinite sequence of magnetization plateaux due to the long-range interaction where each plateau corresponds to a distinct ordered state of matter with a different unit cell.
We calculate the ground-state magnetizations using a classical naive greedy optimization and quantum annealing on the D-Wave Advantage\texttrademark\ system using unit cells up to 36 spins according to the construction in \cite{Koziol2023,Koziol2024,Koziol2025}.
For the considered unit-cell sizes and grid in $h/J$ we find perfect agreement in the devil's staircase between the two methods.

\begin{figure}
	\centering
	\includegraphics[]{"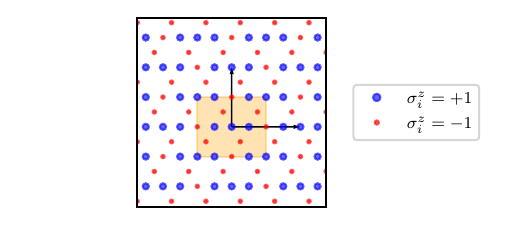"}
	\caption{Magnetic configuration of the candidate ground state of the dipolar ($\alpha=3$) antiferromagnetic LRIM on the Kagomé lattice with no longitudinal field determined with the UCBOS with quantum annealing optimization on the D-Wave Advantage\texttrademark\ system. Blue circles refer to spins with $\sigma^z_i=+1$ and red circles to spins with $\sigma^z_i=-1$. The shaded region corresponds to the unit cell of the ordered structure.}
	\label{fig:kagome}
\end{figure}

To determine the ground state of the dipolar antiferromagnetic LRIM on the Kagomé lattice with no longitudinal field is a long-standing problem motivated by artifical spin ice \cite{Chioar2016,Skjrv2019}.
Especially, the study of the model with the untruncated long-range interaction is of great interest, since from the current literature \cite{Chioar2016} a different ground-state pattern is expected than for models truncated after the third-nearest neighbours \cite{Colbois2022}.
Chioar et al. \cite{Chioar2016} determined a ``very likely choice'' \cite{Chioar2016} for the ground state of the untruncated model using classical Monte Carlo simulations.
They proposed the same state as we find using the UCBOS with quantum annealing and present in Fig.~\ref{fig:kagome}.
With this calculation we confirm the finite temperature classical Monte Carlo calculations \cite{Chioar2016} suffering from freezing by our ground-state optimization protocol.
The interesting aspect of this result is that, when examining a model that truncates the dipolar decay after the third nearest-neighbor interaction \cite{Colbois2022}, one would predict a different class of ground state than when considering the entire tail of the interaction \cite{Chioar2016} (see Fig~\ref{fig:kagome}).
Notably, when considering van-der-Waals interactions $(\alpha=6)$ both truncated \cite{Colbois2022} and untruncated \cite{Koziol2023} studies predict the same class of ground states.

\begin{figure}
	\centering
	\includegraphics[]{"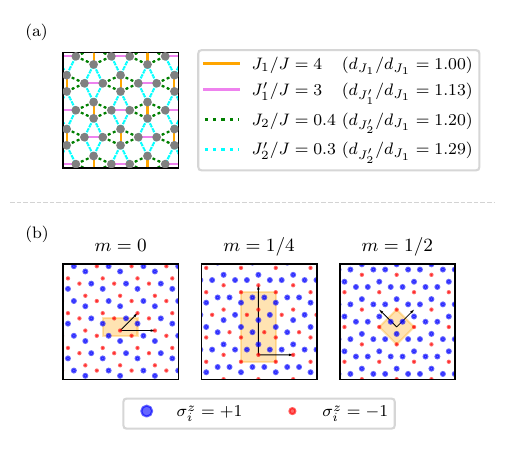"}
	\caption{(a) Illustration of the geometry and additional anisotropic nearest-neighbor Ising couplings in the anisotropic Shastry-Sutherland lattice with realistic Ising coupling values $J_1$, $J_1^{\prime}$, $J_2$, and $J_2^{\prime}$ and interatomic distances $d_{J_1}$, $d_{J_1^{\prime}}$, $d_{J_2}$, and $d_{J_2^{\prime}}$ for the $\text{Er}_2\text{Be}_2\text{Ge}\text{O}_7$ compound \cite{Yadav2025}. $J$ denotes the amplitude of the additional dipolar ($\alpha=3$) Ising interaction. (b) Magnetic configurations at magnetizations $m$ of $0$, $1/4$, and $1/2$ in units of the saturated magnetization determined with the UCBOS with quantum annealing optimization on the D-Wave Advantage\texttrademark\ system for the model with anisotropic nearest-neighbor and dipolar long-range interactions. Blue circles refer to spins with $\sigma^z_i=+1$ and red circles to spins with $\sigma^z_i=-1$. Shaded regions correspond to the unit cells of the ordered structures.}
	\label{fig:SSL}
\end{figure}

In the UCBOS it is possible to add additional few nearest-neighbor interactions, as long as the interactions remain resummable \cite{Dorier2008,Yadav2025}.
Such models become relevant to describe low-temperature properties of quantum materials where the relevant spin degrees of freedom interact with short-range and potential dipolar long-range Ising interactions \cite{Yadav2025}.
The recent study \cite{Yadav2025} considers the compound $\text{Er}_2\text{Be}_2\text{Ge}\text{O}_7$ which is modeled by an Ising model on an anisotropic Shastry-Sutherland lattice \cite{Shastry1981}.
To study the influence of potential dipolar spin interactions in the compound, the UCBOS with a classical optimization was used in Ref.~\cite{Yadav2025}.
We demonstrate in Fig.~\ref{fig:SSL} that for the experimentally relevant magnetization plateaux, we can determine the same ground states of the model with realistic coupling strength using the quantum annealing hardware as with the classical optimizer.
The model used contains additional nearest-neighbor interactions with amplitudes given in Fig.~\ref{fig:SSL} in addition to a dipolar interaction between all atoms.
The resulting magnetization plateaux determined with the quantum annealing optimization are in perfect agreement with the results of the classical optimization provided in Ref.~\cite{Yadav2025}.
This shows that the UCBOS with an optimization using quantum annealing can be directly applied to investigate the relevance of long-range interactions in frustrated materials.

\section{Benefits and Limitations}
It is beyond the scope of this work to derive and probe general statements about the potential advantage of quantum annealing over classical algorithms.
Nevertheless, we would like to report and summarize the observations and learnings we made for the considered problems relevant for the UCBOS.

In general, we observed the expected reduction \cite{Munoz2025} in computation time when comparing computation times of classical optimizations with the time spent on the quantum annealing device for the considered problem sizes.
To calculate the devil's staircase presented in Fig.~\ref{fig:staircase}, we used 10 minutes of processing time \footnote{This includes not only the annealing time, but also the overhead of accessing, programming and measuring the system.} on the quantum annealer and $2500$ minutes ($35$ minutes on $72$ cores) of CPU time on a node with two Intel Icelake 8360Y processors.
We are aware that both computation times could be further optimized and that it is not clear if comparing CPU time of general purpose computer with time used on a purpose build quantum annealer does justice to the matter, but we expect the provided numbers to give an orientation for the capabilities of classical and current quantum devices.
Note, it is not advisable to concatenate too many of the smaller optimization problems to pack them in one go on the quantum annealer since if the problem size becomes too large, larger chains of hardware qubits are needed for the minor embedding which increases the risk of chain breaks.

The current D-Wave Advantage\texttrademark\ system with the Pegasus qubit connectivity advises against embeddings with a chain length larger than $7$ \cite{McGeoch2020}.
The largest unit cell that can be embedded with the all-to-all clique minor embedding algorithm with equal chain lengths fulfilling this condition has therefore $64$ sites \cite{Boothby2020}.

The potentially most interesting point to discuss is up to which unit-cell-size does the quantum annealing determine the optimal state on a unit cell reliably within a reasonable number of annealing runs for the UCBOS application.
As we need to probe many unit cells with the UCBOS and compute time on quantum annealers is limited and precious, it is not feasible to spend arbitrary many annealing runs per parameters value. 
We found heuristically 1000 runs to be a good trade-off for our application case.
An eye catching observation can be made in this regard when looking at the ground state of the dipolar LRIM on the Kagomé lattice.
The state with the lowest energy in the thermodynamic limit in Fig.~\ref{fig:kagome} has a unit cell with 12 sites.
We find this state on its primitive 12-site unit cell, the twice as large 24-site unit cells, but not reliably on the 36-site unit cells (and the even larger ones) that should support this state.
Here, we obtain other states from the annealing, which have a larger energy per site in the thermodynamic limit.
Note, this issue is pattern dependent and it may be that other patterns can be easier determined on larger unit cells since on the triangular lattice we obtain perfect agreement with the classical optimization using unit cells up to 36 spins including states that have a 36-site unit cell.
Also for the magnetization patterns shown in Fig.~\ref{fig:SSL}, we find them consistently also on larger unit cells.
Nevertheless, to be safe, one should be cautious of the annealing results for unit cells with more than 30 sites.

\section{Conclusion}
In the article, we showed a UCBOS that enables the search of ground states of lattice models with algebraically decaying competing long-range interactions on commercial D-Wave Advantage\texttrademark\ quantum annealing devices.
We demonstrated the approach for three different LRIM problems motivated by atomic quantum simulation platforms, artificial spin ice, and the study of frustrated materials with large Ising spins.
Our demonstration suggests that the \mbox{UCBOS} with quantum annealing matches the capabilities of the approach using classical optimization when regarding the results obtained.
While providing the same level of results, the quantum annealing offers a substantial speed-up when comparing with the classical CPU time requirements.
But caution is advised, when relying on the quantum annealing results for larger unit-cell sizes since we heuristically observe that known optimal states are no longer highly likely to be found in a reasonable number of annealing repetitions.

\section{Outlook}
In general, the technical advancements to improve the results of the optimization of the all-to-all connected problems are in line with the known design constraints identified for superconducting qubit quantum annealing devices \cite{Bunyk2014}.
Therefore, we expect advancements in the available devices to transfer directly to improved results of our optimization scheme. 
Also with increasing commercialization and more availability of quantum annealing time, more annealing runs to determine optimal states for larger unit cells will become more feasible.
A future application of the methods could be the study of problems of the form \eqref{eq:LRIM} with larger local Hilbert spaces such as atomic limits of long-range density-density interacting Fermi- or Bose-Hubbard models relevant for quantum dot arrays \cite{Byrnes2008,Buluta2009,Hensgens2017,Knorzer2022}, Moiré materials \cite{Regan2020,Li2021,Huang2021,Nuckolls2024}, or ultracold gases in optical traps \cite{Chomaz2022}.
Note, it is possible to map general quadratic discrete optimization problems onto Ising (binary) problems by using, for example, one-hot bit representations of the discrete states.
This makes these models, in principle, approachable by the quantum annealing architecture, but the number of hardware qubits to represent a logical state of a single lattice site grows at least as the size of the local state space.
Further, to look beyond the horizon of the D-Wave architecture, a more involved concept to omit the issue of increasing chain lengths with increasing unit-cell sizes is the Lechner-Hauke-Zoller code which maps all-to-all connected spin glass problems to a framework with only geometrically local interactions and constraints \cite{Lechner2015}.
In addition to that, this code also serves as a classical repetition error-correcting code \cite{Lechner2015,Pastawski2016}.
So in principle both of these quantum annealing architectures can be used as optimizers for the UCBOS with their strengths and weaknesses.

\section{Acknowledgements}
The authors thank Siddardha Chelluri for fruitful discussions and Anja Langheld and Calvin Krämer for their comments to the manuscript.
The authors gratefully acknowledge the Jülich Supercomputing Centre (\url{https://www.fz-juelich.de/ias/jsc}) for funding this project by providing computing time on the D-Wave Advantage™ System JUPSI through the Jülich UNified Infrastructure for Quantum computing (JUNIQ).
Further, we greatfully acknowledge the scientific support and HPC resources provided by the Erlangen National High Performance Computing Center (NHR@FAU) of the Friedrich-Alexander-Universität Erlangen-Nürnberg (FAU).
The hardware of NHR@FAU is funded by the German Research Foundation (DFG).
This work was funded by the Deutsche Forschungsgemeinschaft (DFG, German Research Foundation), Project-ID 429529648–TRR 306 QuCoLiMa (Quantum Cooperativity of Light and Matter). 
The authors acknowledge the support by the Munich Quantum Valley, which is supported by the Bavarian state government with funds from the Hightech Agenda Bayern Plus.

\textit{Data availability}--
The data that support the findings of this article are openly available \cite{KoziolZenodo}. 
The codes used to generate the numerical results presented in this work can be made available from the authors upon reasonable request.

\section{Methods}

\subsection{Definition of lattices and unit cells}
For the demonstrations of the UCBOS we use the following lattice definitions and unit cells.

\subsubsection{Triangular lattice}
For the evaluation of the devil's staircase on the triangular lattice, we define the triangular lattice as a Bravais lattice with primitive translation vectors $\vec t_1^{\ \triangle}=(1,0)^\text{T}$ and $\vec t_2^{\ \triangle}=(1/2,\sqrt{3}/2)^\text{T}$. To generate the unit cells, we use all distinct unit cells with translation vectors $\vec T_1$ and $\vec T_2$ out of the set
\begin{align}
\label{eq:settriangular}
		\mathcal{A} = &\{i\vec t_1^{\ \triangle}+j\vec t_2^{\ \triangle}|\\ 
        &\nonumber \qquad i\in\{-6,...,6\},\\
        &\nonumber \qquad j\in\{\max(-6-i,-6),...,\min(6-i,6)\}\\
        &\nonumber \} \ .
\end{align}

\subsubsection{Kagomé lattice}
For the evaluation of the ground state of the dipolar antiferromagnetic Ising model on the Kagomé lattice in the absence of a field, we define the Kagomé lattice as a triangular lattice with three sites per elementary unit cell $\vec \delta_1^{\ \text{KGM}}=(0,0)^\text{T}$, $\vec \delta_2^{\ \text{KGM}}=(1/2,\sqrt{3}/2)^\text{T}$, and $\vec \delta_3^{\ \text{KGM}}=(-1/2,\sqrt{3}/2)^\text{T}$ with primitive translation vectors $\vec t_1^{\ \text{KGM}}=(2,0)^\text{T}$ and $\vec t_2^{\ \text{KGM}}=(1,\sqrt{3})^\text{T}$. To generate the unit cells, we use all distinct unit cells with translation vectors $\vec T_1$ and $\vec T_2$ out of the set
\begin{align}
\label{eq:setkagome}
		\mathcal{B} = &\{i\vec t_1^{\ \text{KGM}}+j\vec t_2^{\ \text{KGM}}|\\ 
        &\nonumber \qquad i\in\{-6,...,6\},\\
        &\nonumber \qquad j\in\{\max(-6-i,-6),...,\min(6-i,6)\}\\
        &\nonumber \} \ .
\end{align}
We limit the considered unit cells to a maximium number of 64 sites.

\subsubsection{Anisotropic Shustry-Sutherland lattice}
For the evaluation of the magnetization plateaux of the anisotropic Shustry-Sutherland Ising model with nearest-neighbor and dipolar interactions, we define the anisotropic Shustry-Sutherland lattice as a square lattice with four sites per elementary unit cell 
\begin{align}
 \vec \delta_1^{\ \text{SSL}}&=(0,0)^\text{T} \\
 \vec \delta_2^{\ \text{SSL}}&=(1.094581,1/2)^\text{T} \\
 \vec \delta_3^{\ \text{SSL}}&=(1.094581,-1/2)^\text{T} \\
 \vec \delta_4^{\ \text{SSL}}&=(2.189162,0)^\text{T}
\end{align}
with primitive translation vectors 
\begin{align}
\vec t_1^{\ \text{SSL}}&=(-1.660144,-1.660144)^\text{T}\\
\vec t_2^{\ \text{SSL}}&=(+1.660144,-1.660144)^\text{T} \ .
\end{align}
To generate the unit cells, we use all distinct unit cells with translation vectors $\vec T_1$ and $\vec T_2$ out of the set
\begin{align}
\label{eq:setssl}
\mathcal{C} = \{i\vec t_1^{\ \text{SSL}}+j\vec t_2^{\ \text{SSL}}|i\in\{-6,...,6\},j\in\{-6,...,6\}\} \ .
\end{align}
We limit the considered unit cells to a maximium number of 64 sites.

\subsection{Quantum annealing protocol}

We use the Ocean\texttrademark\ software development kit provided by D-Wave to formulate the optimization problems and address the Advantage\texttrademark\ system quantum annealing device.

As a sampler, we choose the \texttt{DWaveCliqueSampler} class, as recommended for all-to-all connected problems (cliques).

With the \texttt{DWaveCliqueSampler.sample\_ising} method, we communicate the optimization problems to the quantum processing unit with additional parameters \texttt{num\_reads=1000} and \texttt{annealing\_time=200}.
This results in $1000$ annealing runs per optimization problem with an annealing time of $200\,\mu\text{s}$ with the standard annealing protocol.

\end{document}